# Scientific understanding across disciplines: *in varietate concordia?*

## Authors

**Henk W. de Regt**[1], Institute for Science in Society, Radboud University, Nijmegen, The Netherlands

**Milo D. Cornelissen,** Institut für Chemie, Technische Universität Berlin, Straße des 17. Juni 124, TC2, 10629 Berlin, Germany

**Vincent Coumans,** HAN University of Applied Sciences, Nijmegen, The Netherlands

**Nina S. de Boer,** Department of Philosophy, Radboud University, Nijmegen, The Netherlands

**Sebastian De Haro,** Institute for Logic, Language and Computation and Institute of Physics, University of Amsterdam, The Netherlands

**Paul H. W. Disberg,** School of Physics and Astronomy, Monash University, Clayton, Australia / The ARC Center of Excellence for Gravitational Wave Discovery---OzGrav, Australia

**Kristian González Barman,** Centre for Logic and Philosophy of Science, Department of Philosophy and Moral Sciences, Ghent University, Blandijnberg 2, 9000 Ghent, Belgium

**Maarten G. Kleinhans,** Department of Physical Geography and Descartes Centre for the History and Philosophy of the Sciences and the Humanities, Utrecht University, The Netherlands

**Daniel Kostić,** Institute of Philosophy and Sociology, Polish Academy of Sciences, Poland

**Ludo L. J. Schoenmakers,** Konrad Lorenz Institute for Evolution and Cognition Research, Klosterneuburg, Austria

## Abstract

Understanding is a generally acknowledged aim of science, but the way in which understanding is pursued and criteria for when it is achieved differ widely across scientific disciplines. This paper offers a comparative analysis of this disciplinary variation, aimed at revealing differences as well as commonalities in methods and criteria for achieving understanding. Starting point is the contextual theory of scientific understanding (De Regt 2017), which was developed to accommodate contextual variation in criteria for understanding. While this work focused on historical variation and used mainly case studies from the history of physics, the current paper applies the contextual theory to disciplinary variation. We present brief descriptive accounts of how understanding figures in nine scientific disciplines: mathematics, theoretical physics, astrophysics, chemistry, geoscience, biology, neuroscience, psychology, and the engineering sciences. Subsequently, we compare and evaluate these accounts in light of the contextual theory of scientific understanding.

[1] Corresponding author: henk.deregt@ru.nl

## 1. Introduction

Understanding is an aim of most, if not all, sciences. A look at scientific practice teaches us, however, that criteria for when understanding is achieved may differ widely across scientific disciplines. In the philosophy of science this is first of all visible in the debate on scientific explanation, the royal road to scientific understanding. While traditionally philosophers searched for a universal (e.g. nomological or causal) account of explanation, the advent of philosophies of the special sciences went hand in hand with a diversification of models. Causal explanation, for example, is still regarded as a mode of explanation with a broad scope, but nowadays precise articulations of it are more sensitive to disciplinary variation. Moreover, different forms of non-casual explanation have been proposed as being characteristic of the life and social sciences (e.g. functional, topological, and narrative explanation). These observations invite a comparative analysis of the ways in which understanding is achieved in different scientific disciplines: what is shared and which differences can be found? Given the diversification of modes of explanation and criteria for understanding, can we still identify a common core? The present paper will address these questions.

Starting-point is the contextual theory of scientific understanding presented in *Understanding Scientific Understanding* (De Regt 2017), which has been developed to accommodate contextual variation in criteria for understanding. However, the analysis on which this theory is based focuses on variation of historical contexts rather than on disciplinary variation, and uses mainly case studies from the history of physics. Yet, the theory allows for disciplinary variation as well, and hence may have a wider domain of application than physics or the physical sciences. Our paper explores the disciplinary variation of ways in which understanding is achieved, as well as their commonalities.

A core tenet of the contextual theory of scientific understanding is the idea that scientific theories need to be intelligible, where standards of intelligibility may vary with context. Its focus on physical sciences led to special attention for intelligibility criteria that are suited to mathematized theories, which are the rule in this domain. In other disciplines, however, the nature and role of theories may differ, and it remains to be seen how the intelligibility condition applies there, if at all. Our paper addresses this issue via a comparative analysis of criteria for understanding and intelligibility standards in a variety of scientific disciplines, ranging from theoretical physics to psychology. Accordingly, the main aim of the paper is to examine whether, or to what extent, the ways in which scientists acquire understanding in disciplines other than physics can be accounted for by the contextual theory. A secondary aim is to investigate how the contextual theory could be modified in order to provide a more general account of understanding that applies to disciplines beyond physics.

The outline of the paper is as follows. In Section 2 the contextual theory of scientific understanding is summarized. Section 3 presents brief descriptive accounts of how understanding figures in nine disciplines: theoretical physics, astrophysics, chemistry, geoscience, neuroscience, biology, psychology, the engineering sciences, and mathematics. These accounts focus on concepts that play key roles in the contextual theory: explanation, theory, model, intelligibility and skills. Finally, in Section 4, we compare and evaluate the ways in which understanding is produced and the criteria according to which it is assessed in the various disciplines, and interpret the differences and commonalities in light of the contextual theory of scientific understanding.

## 2. The contextual theory of scientific understanding

The central idea of contextual theory of scientific understanding is the thesis that scientists have achieved understanding of a phenomenon *P* if they possess an explanation of *P* on the basis of an intelligible theory *T*. This is summarized in a Criterion for Understanding Phenomena (De Regt 2017, 92):

> **CUP**: A phenomenon *P* is understood scientifically if and only if there is an explanation of *P* that is based on an intelligible theory *T* and conforms to the basic epistemic values of empirical adequacy and internal consistency.

The key term in this criterion is 'intelligible', where intelligibility is defined as (De Regt 2017, 40):

> **Intelligibility**: the value that scientists attribute to the cluster of qualities of a theory *T* (in one or more of its representations) that facilitate the use of *T*.

Note that **CUP** entails that achieving scientific understanding not only requires intelligibility but also additional values: empirical adequacy and internal consistency. Explanatory models must adhere to these epistemic values to be considered scientific. Following Kuhn, these are referred to as values rather than strict criteria, as their importance can be ranked and applied differently depending on the context (see De Regt 2020).

Intelligibility is not an intrinsic property of theories, but a context-dependent value: whether or not a theory is intelligible to scientists depends on, for example, their skills and their background knowledge. Why do scientific theories need to be intelligible to the scientists who use them? The argument for this claim draws on the work of Cartwright (1983, 143-162) and Morgan and Morrison (1999), who highlighted the pivotal role of modelling in scientific practice, and in explanatory practices in particular. On this model-based account of scientific explanation scientists acquire understanding of the phenomena by constructing models, which 'mediate' between relevant theories and the phenomenon-to-be-explained (Morgan and Morrison 1999). Constructing such mediating models involves pragmatic judgments and decisions, since models do not follow straightforwardly from theories (and neither do they follow from the empirical data). For example, suitable idealizations and approximations need to be made. Hence, the construction of models – which provide explanatory understanding of phenomena – requires theories that are intelligible in the sense defined above. Only if scientists' ability to work with the theory allows them to make suitable pragmatic judgments, will they succeed in constructing explanatory models. In sum, understanding a phenomenon on the basis of *T* depends on an appropriate combination of skills of *S* and qualities of *T*. Preferred theoretical qualities can be regarded as 'tools for rendering theories intelligible'.

An example of such a quality is visualizability. This is a theoretical quality that is widely valued, because, for many scientists, visualizable theories are more tractable and easier to work with (De Regt 2017, 226-259). But the contextual theory does not imply that visualizability is a necessary condition for the intelligibility of scientific theories. Depending on the context, there may be alternative ways to render theories intelligible. Whether a theory is intelligible to scientists – that is, whether they value specific qualities of the theory – depends on contextually varying factors, most notably on their skills. It seems plausible that both the historical and the disciplinary context play key roles in this respect.

One might worry that the context-relativity of intelligibility, and its value-laden nature, renders it devoid of normative power. If valuing a theory as intelligible is merely a matter having a subjective preference for that theory, it may seem that there can be no objective (normative) criteria for assessing whether a theory is intelligible. However, this does not follow. Although intelligibility criteria are contextual and will vary across scientific communities (and sometimes even within a community), there are ways of objectively testing whether a theory is intelligible to scientists in a specific context. Such an objective test is provided by the following Criterion for the Intelligibility of Theories (De Regt 2017, 102):

> **$CIT_1$**: A scientific theory T (in one or more of its representations) is intelligible for scientists (in context C) if they can recognize qualitatively characteristic consequences of T without performing exact calculations.

The basic idea behind $CIT_1$ is that its fulfilment indicates that scientists have an 'insight' into the workings of the theory and are accordingly able to use the theory for the construction of models of the phenomena. If the resulting models are empirically adequate and internally consistent, they provide understanding of the phenomena in question. The conceptual tools – e.g. visualization – that scientists may use to recognize qualitative consequences of the theory are also the tools that will facilitate model construction. $CIT_1$ is only one of many possible criteria (tests) for intelligibility: it is a sufficient condition for intelligibility, not a necessary one. As it applies specifically to quantitative theories, it appears useful mainly in the physical sciences, where such theories are the rule. In other disciplines other criteria may (or should) be employed; hence the subscript: it is one version of CIT, others might be $CIT_2$, $CIT_3$, etc. In the next section we will investigate how standards of intelligibility vary across disciplines, and whether criteria CUP and $CIT_1$ remain applicable in a wider domain than the physical sciences.

## 3. Understanding in different scientific disciplines

This section presents brief accounts of methods and criteria for achieving understanding in nine scientific disciplines: theoretical physics, astrophysics, chemistry, geoscience, biology, neuroscience, psychology, the engineering sciences, and mathematics. As it is obviously impossible to present exhaustive analyses of how scientific understanding is generated in these disciplines, the aim of this section is to highlight practices that can be accounted for by the contextual theory of understanding. If successful, our examination shows that these disciplines can be accommodated at least partially with the contextual theory. In order to make a comparison with the contextual theory possible, the accounts in the subsections focus on key concepts in the contextual theory of scientific understanding: explanation, theory, model, intelligibility and skills. Questions that will be addressed are:

- Explanations: Which forms of explanation are employed?
- Theories and models: What is their nature and function?
- Intelligibility standards: Which theoretical qualities are valued?
- Skills: Which skills do scientists need and use for understanding?

While all subsections answer these four questions for their specific discipline, their structure varies because the relevant literature and the state of philosophical reflection differ across disciplines. In fact, this variation in the literature plausibly reflects disciplinary differences in how understanding has been theorized and can thereby be seen as a first indication of disciplinary variation.

### Theoretical physics

Theoretical physics is a broad area, comprising a variety of fields and methodologies. All major types of explanation can therefore be found in the practice of theoretical physics: especially model-based, deductive-nomological, causal and mechanistic explanations. Theoretical physicists build models of phenomena through the application of theory. They also use models as *starting points* for further theory development. Both mathematical and iconic or representational models are used. For example, the Georgi-Glashow (1974) model has often been used to study low-energy solutions of quantum field theories, because it has some mathematically simpler properties than the standard model of particle physics and illustrates some of its qualitative behaviours, such as the mechanism for the generation of mass by the Higgs field. Early models of quark confinement used a mix of visual representations and mathematics to model the confinement of colour charge inside of subatomic particles, on an analogy with the Meissner effect in type II superconductivity, where the magnetic field in a superconductor is confined to the interior of vortices. These models used both iconic and mathematical analogies between condensed matter theory and quantum field theory to give an explanation of the confinement of colour charge (De Haro and Butterfield 2025, Chapter 6).

In theoretical physics, an important aim is to transform heuristic and imprecise models into rigorous mathematical models that are (approximate) solutions of a theory (in both examples mentioned above, the standard model of particle physics). Indeed, the prediction of new phenomena, such as the Higgs mechanism and the production of the top quark, is seen as the golden standard of theory assessment. Thus deductive-nomological explanation plays an important role, especially in the more advanced stages of theory development.

The various models of explanation are not mutually exclusive, and they often complement each other. There are three reasons for this: first, as we already mentioned, different types of explanations often correspond to different stages of theory development and mathematical sophistication. Second, although explanations that use mathematical models and deductive-mathematical reasoning are more quantitatively precise, heuristic reasoning through iconic models or visual aids are often also used in explaining, and so remain relevant for introducing sophisticated ideas to newcomers, and for communication with broad audiences. For example, although Hawking's heuristic explanation of black hole radiation in terms of the pair creation of particles is not very rigorous, it offers an intuitive picture that is heuristically fruitful. Third, physicists are usually not satisfied with merely mathematical derivations of phenomena. For it is generally agreed that a derivation, if it is to be a good explanation, must be interpreted in physical language. It is regarded as a virtue that the interpretation can lay bare concrete physical *causes* and *mechanisms*.

There are manifold qualities that make theories intelligible, and tools developed that latch onto those qualities. Broadly speaking, we can distinguish between qualities related to (A) the mathematical formalism of a theory, and (B) its interpretation. (Note that these mathematical and interpretative aspects are often combined.)

The mathematical tractability of a theory is a key quality for understanding how a theory works and how it makes predictions. It is attuned to scientists' skills to apply mathematics in physics. Thus physicists develop mathematical tools that make a difficult problem easier, according to the mathematical skills and standards in the field. Examples include symmetries: identifying a symmetry often makes a difficult problem mathematically tractable. This generalizes to the idea of duality, or isomorphism between theories. Dualities are very powerful tools for theory construction, because they allow the development of both formal and interpretative analogies and, in some cases, even equivalence, between very different theories (De Haro and Butterfield 2025). In other cases, physicists use approximative relations and similarities between theories. Similarities can be both formal and interpretative. For example, the similarities between the conservation laws of different theories can be exploited for rendering a theory intelligible. Thus, the formal similarity between the law of charge conservation in the Maxwell theory and the conservation law for the probability density in quantum mechanics was historically valuable in developing the interpretation of the wave-function (De Haro and De Regt 2018, 2020). Other criteria include consistency and analogy.

Interpretative qualities of theories include the applicability of physical mechanisms and causal connections (often in the context of approximative or idealized models), and whether they admit visualization. Indeed, visualization and, more generally, geometric reasoning, are important tools in theoretical physics. The use of Feynman diagrams in quantum field theory, Penrose diagrams in general relativity, and quantum computing circuits in quantum gravity, attests to this fruitfulness.

## Astrophysics

Astrophysicists study a wide variety of cosmic objects, such as planets, stars, black holes, nebulae, and galaxies. However, it is not straightforward to investigate these objects and their evolution: they often evolve on timescales of millions or even billions of years, and are difficult to observe due to the signals (i.e., light, particles, or gravitational waves) being affected by the large distances and cosmic environments. This means that in order to analyze the evolution of stars or galaxies, astrophysicists need to consider vast numbers of observed objects and extract patterns that describe their evolutionary history. For example, a single star typically evolves on timescales too long to observe, but a stellar population contains stars at various stages of their evolution, based on which evolutionary patterns can be extracted (e.g., Gaia Collaboration 2018). Due to this tracing back of evolutionary histories, astrophysics can be considered a "historical science" as opposed to an experimental one (Cleland 2002; see also Anderl 2016).

In an attempt to formulate explanations of these observed patterns, astrophysicists might make use of computer simulations (i.e., simulations of astrophysical phenomena based on known physics). When simulating a single star, for instance, physical theories dictate the relationships between quantities such as temperature, density, and pressure. These relationships are then used to formulate a model that includes certain assumptions and idealizations concerning the structure of the star, which enables the scientist to numerically integrate the relations and describe, for example, how aspects such as the radius or luminosity evolve over time. The explanatory power of these simulations comes from the fact that they can provide insight into the relationship between the theoretical model and observable quantities of the simulated objects (such as their luminosity).

However, usually these causal relationships do not follow straightforwardly from the simulation. Astrophysical models, after all, can be relatively complicated, because of which it may not be obvious how certain aspects of the theoretical model have shaped the simulated results. Understanding the simulation, therefore, entails having insight into this relationship between theoretical model and simulated results. This is well described by the contextual theory: a simulation is intelligible to astrophysicists if they can recognize its "qualitatively characteristic consequences". For some scientists, this kind of recognition might occur through visually imagining the evolution of the system and then "seeing" how the theoretical model evolves over time (see Disberg 2025, for a more elaborate discussion). Theoretical models might therefore be considered insightful and valuable when it is relatively easy to "see" their qualitatively characteristic consequences, which provide the link between physical theory and (potential) observation. For other scientists, however, it can be more important to show an elaborate mathematical derivation that describes the characteristic consequences of a model, meaning that there is a variety of skills (e.g., imaginative or mathematical) that are useful for astrophysical understanding.

Astrophysical explanation is often based on such a connection between some sort of theoretical model (either simulated or analytical) and observations.[2] If a model is able to reproduce an observed signal, it becomes plausible that the signal is caused by a system similar to the one described by the model. For example, Price et al. (2024) simulated a star being torn apart due to its orbit around a massive black hole, in an attempt to explain why some signals linked to such events (i.e., tidal disruption events) contain optical and ultraviolet radiation. Their model made several assumptions, for example about the structure of the star and about how the gas separated from the star produces light, where they used the theory of general relativity to determine the motion of the gas. Their calculations predict the formation of a bubble of gas around the disrupted star, where the surface of this bubble has a temperature corresponding to emission of light in optical and ultraviolet wavelengths, effectively explaining the observed signal by reconciling theory and observation. Understanding the phenomenon of the observed signal, then, is achieved when scientists can "trace back" the causal chain of this reconciliation (cf. "historical science"), for example through a simulation of which they grasp the characteristic consequences. This way, they gain insight in how a system as described by their model might cause a signal as observed.

[2] Cf. the phrase "saving the phenomena" (see e.g., Hacking 1989; Disberg 2025).

In summary, astrophysics is a 'historical' science in which explanations are usually based on an agreement between observation and theory (e.g., as underlying a constructed model such as a simulation). Understanding a theoretical simulation can involve grasping the qualitatively characteristic consequences of the model. When the simulated results are similar to an observed signal this holds explanatory power and allows for understanding of the phenomenon according to the definition of CUP, given that the simulation---or alternative theoretical analysis---is intelligible.

### Chemistry

Chemical understanding is in many aspects comparable to physical understanding, since chemists explain fundamental properties of molecules such as enthalpy, absorbance, or viscosity with causal and mechanistic theories firmly grounded in physics. The difference, however, lies in chemical phenomena such as solubility or stability, which are not understood quantitatively and tend to lose meaning once mathematized (Hoffmann 1998). By identifying common patterns in structural formulae, chemists can predict under which conditions a molecule is stable and can be transformed in a desired way. How do chemists make such inferences from the composition of a molecule and predict its stability and solubility?

Understanding of stability partly depends on physics, but the most fundamental quantum-mechanical theories are not of much use, since exact solutions are intractable for all but the tiniest of molecules. Theoretical chemists have therefore developed approximations such as molecular orbitals, which represent the regions where electrons have a high probability to be located. This visualization of electrons helps to understand bonding within molecules and allows chemists to understand their reactivity without performing exact calculations ($CIT_1$). Common types of reactions are explained by orbital overlap, wherein a new bond is formed if two molecules can position themselves in such a way that these regions can be overlaid. This allows for qualitative understanding when the orientation of the two molecules required to form a bond can be visualized: obstruction through opposing charges or other types of clashing would make it unlikely for the two molecules to adopt the required conformation, whereas positive effects such as bond angle compression can increase the likelihood of bond formation.

Additional insights are gained through qualitative chemical theories and concepts, which share qualities of visualizability and similarly yield causal and mechanistic explanations. In a case study on synthetic chemistry, two of us have shown how such idealized concepts are used to acquire chemical understanding (Cornelissen and De Regt 2022). As fruitfulness is generally valued over accuracy, these theories typically describe trends among classes of molecules, aiming to predict the properties of compounds relative to one another; the concepts of functional groups and substituent effects, for instance, help to identify common elements within structural formulae, categorizing them to understand the behavior of the different parts of a molecule. Rather than determining the nucleophilicity or acidity of a structural element (e.g. an amino group) for each new molecule, the chemist instead predicts its behavior through a summation of structural patterns previously encountered in other molecules. Further guided by concepts such as induction and mesomerism, explaining the relative ability of these molecules to harbor charges, as well as steric hindrance and insights from conformational analysis, it becomes possible to predict the properties of a certain functional group based on the surrounding structural motifs. By combining the attributes of these common elements, the behavior of new molecules can be understood without calculations for each specific case. Theories that make whole classes of molecules intelligible without the need for calculation, hence possessing $CIT_1$, are preferred and most used.

A clear difference with physics is that many conclusions and explanations in chemistry do not take a mathematical form. Examples of such qualitative reasoning are found in explanations of solvent effects in chemical reactions, for instance in a recent review on palladium cross-coupling reactions (Sherwood et al. 2019). These analyses contain hardly any equations, instead depending on qualitative concepts such as coordination and polarizability to explain catalyst solubility and metal aggregation in different solvents. As Hoffmann (2007) and others have pointed out, chemists often make use of conflicting and even mutually exclusive theories, using this “piecewise understanding” to their advantage. Drawing from this pool of theories, they set up confined experiments and tweak the properties of the mixture to probe the effect of factors such as pH, light, temperature, solvent, or the presence of impurities. Gaining understanding from different angles and on different levels, the result is a back-and-forth series of experiments that slowly reveals the nature of a mixture or reaction. In sum, the chemist is able to switch rapidly from conflicting idealized models and test isolated properties, making parts of the system intelligible to gain overall understanding.

### Geoscience

Geoscience aims for historical descriptions and causal explanations of the structure of the Earth and its development. Some subdisciplines study the physical, chemical and biological mechanisms that, in interaction, cause phenomena on scales from the microscopic to the Solar System. Other subdisciplines reconstruct the development of such phenomena on scales from milliseconds to billions of years. As such, geoscience exhibits aspects of natural and historical sciences (Kleinhans et al. 2005; Bokulich 2021)

We use the familiar phenomenon ‘river delta’ to illustrate the pluralist scientific practices and roles of geoscientific understanding. Deltas form on Earth and on Mars where rivers end at the coast by sedimentation of gravel, sand and mud. On maps and remote sensing imagery, some deltas show active river branches with overlapping lifespans, connected at an upstream channel bifurcation, whereas other deltas are formed by rivers shifting more rapidly from one direction to another by carving a new channel and leaving an abandoned river deposit. Evidence of multi-branching and displacing rivers is amply found in outcrops and subsurface deposits of rivers that were active up to four billion years ago. Geomorphologists and civil engineers study the existence and (in)stability of two river branches connected at a bifurcation, while, independently, geologists study how river deposits were stacked up in the past by river displacement (called avulsion) (Kleinhans et al. 2012).

Geomorphological theories explaining river bifurcation stability are based on the physics of fluid dynamics and sediment motion and known conditions. The (in)stability of a nearly symmetrical channel bifurcation can be understood as the effect of a (dis)balance between a positive and a negative feedback. This theory is consistent with theory that explains the existence of river bars (Kleinhans et al. 2023). The theory explains which specific causal variable ranges determine whether bifurcations return to exact symmetry or develop to the alternative stable system state of extreme asymmetry, wherein one downstream channel receives an increasing proportion of water and sediment until the other is (nearly) abandoned (Kleinhans et al. 2012).

Geomorphological theories have been used to develop mathematical models of suitably simplified physical equations solved analytically, and numerical simulation models with greater complexity and freedom to include external causes of unbalancing the system. Such modelling based on physics is an important skill of geomorphologists. River engineering works in the 18th century to stabilize river bifurcations attest to intelligibility ($CIT_1$) of theory for river erosion and sedimentation (Kleinhans and Van Besouw 2024).

Geological theories identify causes of avulsion and of patterns of observed river deposits in time and space, such as compaction and sedimentation of the floodplain outside the channels, sea-level rise and coastline progradation. Local compaction of clayey and peaty floodplain under the weight of new channel deposits, which attracts more channels like children who naturally cluster together when jumping on an inflatable mattress or trampoline. These causes all lead to more elevated channels and flood water levels that can cause a river to spill over and avulse into the relatively lower floodplain (Mackin 1948, Kleinhans et al. 2012). These causes are qualitatively intelligible in that a larger elevation difference between a river and its surrounding plain makes it more likely that the river in a flood flows down the steeper slope out of the channel onto the plain, which is the start of an avulsion (Paola 2000). An important skill of geologists is reconstruction of past conditions.

Geological theories have been used to create simulation models and analogue, material models of deltas with moving water and sand in the laboratory. These models produce channel deposit geometries, stacking patterns and statistical descriptions that help to infer the formative conditions in outcrops and the subsurface. Visualized model output, sketched conceptual models, and geometric considerations are used to make the theory intelligible ($CIT_1$).

Early in the 21st century, it was recognized that an unstable bifurcation is an important aspect of the avulsion process (Kleinhans et al. 2012). This recognition was possible because some geological studies of avulsed river deposits concerned well-known historic cases of unstable bifurcations. However, the subdisciplines remain largely focused on different questions, spatiotemporal scales and observation techniques, which are not intelligible without different background knowledge and skills, so that there is no general common understanding of bifurcation and avulsion theories. In conclusion, scientific understanding is linked to, and limited by, different skills and background knowledge of the different earth-scientific disciplines.

### Neuroscience

One of the distinguishing features of physics is that it is organized around a set of fundamental theories, such as Newtonian mechanics, electromagnetism, thermodynamics, and statistical mechanics—each of which describes different aspects and scales of physical reality. Neuroscience, by contrast, does not possess such fundamental theories. This is due to the exceptional complexity of the brain, which resists unification under a single theoretical framework. As a result, neuroscience is characterized by a proliferation of modeling and explanatory approaches that are epistemically diverse and often irreducible to one another. Gold and Roskies (2008) famously describe the field as theory-poor but model-rich.

This difference matters. The concept of intelligibility introduced in Section 2 above is framed primarily in relation to theory-driven disciplines like physics, and it is not immediately clear how—or whether—it can be applied to model-based sciences like neuroscience. However, we submit that this does not undermine the contextual theory of understanding, but rather enriches it by pushing toward an even broader conception of intelligibility. This becomes especially clear when we consider how neuroscientists construct and apply models. Rather than deriving these models from unifying theories, they often adapt existing theorems from different areas of mathematics to model neural behaviors or properties. For instance, they use non-linear differential equations to study the dynamics of bi-manual coordination, network control theory to examine cognitive controllability, and graph theory to understand the metabolic costs of neuroanatomical connectivity. These models provide different kinds of explanations, each with their own set of epistemic and ontic commitments. Kostić and Halffman (2023) provide empirical data about the distribution of these explanatory practices in neuroscience. Because these theorems are highly abstract and developed independently of neuroscience, and can therefore be applied in building models in many other areas of science, including physics, their relevance to biological phenomena is rarely straightforward (Kostić 2020; 2024; Woodward 2025). As a result, scientists must rely on various forms of qualitative reasoning to render them intelligible.

For example, the general phenomenon of brain robustness, i.e., its ability to preserve function despite structural-anatomical changes due to lesions or injury, can be studied using network models. In these models, robustness is mathematically understood as invariance under random perturbations. But since such invariance can be mathematically multiply realized, scientists must engage in extensive qualitative reasoning to interpret the models in each specific context of application. A well-known example is the graph-theoretic concept of "small-worldliness," used to model and explain various phenomena in the brain, such as efficient signal processing (Watts and Strogatz 1998), wiring minimization (Stiso and Bassett 2018), and the routing or the navigation problem in brain networks (Seguin et al. 2018). The problem is that the mathematical property of small-worldliness can be instantiated in models in numerous ways, i.e., via densely interconnected local triplets of nodes and a few long-range connections between distant neighborhoods of nodes (Watts and Strogatz 1998), a nested hierarchy of modules of densely interconnected nodes (Hilgetag and Goulas 2016), or a particular statistical distribution of long-range connections in a small-world network (Seguin et al. 2018). Each one of these different ways in which small-worldliness is mathematically instantiated, requires a different qualitative interpretation even when applied in the same context. This is actually desirable because different mathematical instantiations, even of the same abstract property such as small-worldliness, presumably help to accomplish different epistemic goals, which increases our understanding of the phenomenon. Naturally, these concepts, models, and explanatory practices involve a range of skills, including counterfactual, modal, deductive, inductive, abductive, and erotetic reasoning, as well as pattern recognition, data compression, category learning, and concept grasping across both linguistic and mathematical domains. In the examples above, pattern recognition is employed in identifying recurring structural features across network representations, such as densely connected clusters, hierarchical modules, or characteristic distributions of long-range connections; data compression reduces these complex structures to tractable variables or labels, such as clustering coefficients or path lengths; and category learning groups these different mathematical structures under the abstract category of small-world networks. Abductive reasoning supports the inference that a given network instantiates small-worldliness because this hypothesis best explains the observed combination of features, while grasping why different instantiations yield different explanatory patterns involves counterfactual and modal reasoning. These skills can themselves be studied empirically using tools from the naturalized epistemology of understanding (Khalifa et al. 2022).

To accommodate the shift of focus from theories to models in neuroscience, and also as part of their broader idea of what they call the "rich interpretation" of scientific models, Kostić and Khalifa (2025) suggest a revision of criterion $CIT_1$:

> **CIR**: scientific representation *R* is intelligible for scientists (in context *C*) if they can recognize the qualitatively characteristic consequences of *R* without performing exact calculations.

Initial intuitions notwithstanding, neuroscience challenges and enriches the concept of intelligibility by exemplifying a theory-poor, model-rich science that relies on highly abstract, context-sensitive representations and qualitative reasoning. Crucially, it shifts the focus of intelligibility from scientific theories toward scientific models.[3] This motivates a partial revision of the contextual theory and anchors intelligibility more firmly in scientific practice, by showing how neuroscience affords empirically tractable cases for studying the computational and biological underpinnings of diverse forms of qualitative reasoning.

### Biology

Biology is a deeply heterogenous discipline. Its content and disciplinary boundaries are difficult to define beyond each its constituent disciplines having to do with life in some relevant way – a concept that is itself notoriously difficult to define (Schoenmakers 2023). Much like life itself, we know biology when we see it, yet the discipline as a whole is highly disunified (Leonelli 2009). The biological sciences encompass disciplines as diverse as ecology, evolutionary biology, mycology, anatomy, cell biology, botany, systems biology, behavioural ecology (or ethology), botany, and microbiology, and they include boundary disciplines with various other sciences, such as biochemistry, biophysics, and bioinformatics. The practitioners of each of these biological sciences show enormous overlap as well as tremendous differences in the objects that they study, the methodological approaches that they employ, the scientific theories and modelling approaches that they rely on, the epistemic values and metaphysical commitments that they hold dear, and the standards of adequacy that they insist upon. The biological sciences, perhaps more so than any other scientific discipline, show their unity in diversity.

This diversity in disciplinary and epistemic practices makes it *prima facie* unlikely that CUP and $CIT_1$ can be applied across the biological sciences in the same way as they can be applied in physics. One reason is that CUP and $CIT_1$ are focused on the intelligibility of scientific theories, while large parts of the biological sciences are model-focused rather than theory-focused. In this sense, biology is similar to neuroscience and $CIT_1$ could be exchanged for CIR, on which biological understanding results from recognizing qualitatively characteristic consequences of scientific representations (rather than theories) without performing exact calculations (Kostić & Khalifa, 2025; see above). For biology, the shift from theories to representations is a step in the right direction, but CIR too is unlikely to cover all of the biological sciences. Just as the importance of theories varies across the biological sciences, so do the meaning and importance of exact calculations. To give only one example, exact calculations (beyond basic arithmetic) often have little bearing on natural history, which remains a key source of understanding in fields such as behavioural ecology (Endler 2025; Guevara-Fiore 2025; Travis 2020).

Nevertheless, we submit that the contextual theory of scientific understanding can accommodate the diversity found in the biological sciences. Recent investigations into the role of understanding in the biological sciences support this claim and reveal several important insights. First, scientific understanding is seen as a key aim of the biological sciences, ranging from disciplines that rely heavily on mathematical modelling, such as models for foraging (Rice 2016) and cooperation (Potochnik 2017) in behavioural ecology or the biophysical description of nerve cell propagation (Holland et al. 2024), to the disciplinary integration of evolution and development (Cortés-García and Lopez-Orellana 2019). Second, and similarly to physics, visualizability is of crucial importance in conferring understanding in biological disciplines such as ecology (Poliseli 2020) and computational systems biology (Stuart and Nersessian 2019). Third, the grasping of mechanistic explanations is particularly important for understanding in the biomedical sciences (Varga 2024), where it serves both a scientific and a clinical purpose, and in ecology (Poliseli 2020), where visualization has been identified as an important mediator of mechanistic understanding. Fourth and finally, beside theoretical skills (e.g., working with theories, models, or conceptual frameworks), the biological sciences typically involve a large and variable number of performative skills (e.g., working with certain equipment, specimens, or within certain physical environments) and social skills (e.g., working according to the communicative standards of the field, or working within certain social environments) that form part of the pragmatics of understanding (Leonelli, 2009). Importantly, these skills may not always be easy to identify through purely historical, retrospective analyses of how scientific understanding was achieved, but require supplementing history and philosophy of science with ethnographic approaches (Poliseli 2020).

---

[3] This still accords with the contextual theory's central thesis that it is the model that produces explanatory understanding. Incidentally, this is not to deny that models can serve other epistemic goals as well, for example by providing exploratory heuristics for identifying relevant parts or features of a system that are not accessible using other kinds of models (Serban 2020; Rivelli 2019). Canonical examples in the literature include mechanistic and topological models. In neuroscience, network models are sometimes used to identify relevant components of a causal structure, thereby contributing to the completion of a mechanistic model. Conversely, dynamical models are often employed to add detail to highly abstract topological models. For instance, a topological model may represent global mathematical constraints on anatomical connectivity in the brain, but without an accompanying dynamical model it remains unclear how signals propagate through anatomical networks (Woodward 2025). These modeling practices also rely on qualitative reasoning to support further inferences about neural structure and dynamics, to generate new research questions, and more generally to guide scientific inquiry.

The upshot is that, despite the diversity found in the biological sciences, understanding plays an important role in biology, and biology can be accommodated within the contextual theory of scientific understanding. Similar to neuroscience, understanding in biology requires a broader conception of the intelligibility than captured in $CIT_1$, however, rather than replacing $CIT_1$ with CIR, we expect that this broader conception involves a number of CI's that cover the diversity in disciplinary and epistemic practices found in the biological sciences. Investigating the role of understanding within a greater range of the biological sciences, and formulating the accompanying criteria for intelligibility, is a significant but crucial task for a fuller and more general understanding scientific understanding in biology. The biological disciplines that have been examined thus far show that this project is feasible and appropriate.

### Psychology

Psychological science is an umbrella term for disciplines that scientifically study the human mind and human behavior, ranging from developmental and cognitive psychology to social and cross-cultural psychology. Psychological science incorporates a broad spectrum of research methodologies. Here, we use the term to refer to *quantitative* psychological science whose primary output is statistical patterns, e.g., mean group differences, partial correlations, and statistical interactions. Psychological scientists conduct research to obtain, among other things, a scientific understanding of psychological phenomena. To date, however, the details of such understanding, in psychological science in general, or its subdisciplines, have rarely been explicated (but see De Boer 2025; Eigner 2009; 2010). Given its focus on statistical patterns, much of psychological science involves the construction of and interaction with *data models*. These data models are representations that summarize and describe quantitative data on psychological phenomena gathered via measurements or observations. Examples of these are latent variable models or network models estimated via network analysis. Although building and interpreting these data models could foster understanding, they are best described as *exploratory*. This means that their understanding-providing potential may not be well-captured by the contextual theory. Instead, we zoom in on the *explanatory* understanding provided by psychological theories.

Psychological science aims to provide a range of explanations of psychological phenomena, including mechanistic, causal-interventionist, functional, and non-causal, dynamical explanations. Indeed, recent accounts have advocated for explanatory pluralism in (subdisciplines of) psychological science (e.g., Potochnik and Sanches de Oliveira 2020). Which explanatory strategy is preferred may differ per subdiscipline. For instance, clinical psychological science studies phenomena such as the vulnerability of specific demographics to developing mental problems and the efficacy of psychological treatments. It does so with the – sometimes distal – aim of *reducing* mental suffering, demonstrating a preference for counterfactual and, ideally, causal explanatory strategies.

However, the status of these explanations differs from those in the physical sciences. To date, most psychological theories are qualitative and yield unclear predictions (Meehl 1978). Psychological scientists have proposed frameworks to advance the construction of psychological theory, for instance, by using formal modeling (Borsboom et al. 2021; Fried 2020; Haslbeck et al. 2022; Van Dongen et al. 2024). However, theory construction in psychological science is complicated by the lack of robustness and construct validity of many psychological phenomena, and the difficulty in identifying difference-making relationships between psychological variables (Eronen and Bringmann 2021). These difficulties may be more pronounced for specific subdisciplines, such as clinical psychological science, whose phenomena of interest include descriptive and evaluative dimensions and are subject to looping effects. So, psychological theories provide *how-possibly* explanations; applying CUP to psychological science would thus require broadening the criterion to account for such explanations.

Psychological scientists value different qualities in psychological theories (and theoretical models). First, they value theories with explanatory precision (i.e., explanations for precise psychological phenomena) and explanatory breadth (i.e., explanations for a broad range of psychological phenomena) (Borsboom et al. 2021; Haslbeck et al. 2022). Second, they value qualities that improve scientists' ability to use and reason with these theories, i.e., qualities that make them more intelligible. For instance, Borsboom et al. (2021) value psychological theories that demonstrate simplicity and draw on analogies. Moreover, in line with the notion of *Verstehen*, psychological scientists may value theories that align with folk psychological intuitions about our reasons for acting, although the necessity – and even value – of this theoretical quality is contested (e.g., Van Dongen et al. 2024). Finally, data models in psychological science are praised for their visualizability and transparency (Borsboom 2022); these qualities may also apply to (the intelligibility of) psychological theories.

Psychological scientists are primarily trained in the skills needed to construct and interact with data models, such as setting up experimental designs and conducting null-hypothesis statistical testing. Understanding psychological theories may require additional reasoning-related skills. Some call for training in computational modelling to aid psychological scientists in theory construction (Borsboom et al. 2021; Van Dongen et al. 2024). Constructing computational models could improve the intelligibility of psychological theories by providing ‘thinking tools’ that help psychological scientists reason about possible consequences of their theories. Such intelligibility could also be achieved by improving psychological scientists’ skills in metaphorical (Eigner 2010) or analogical reasoning (Borsboom et al. 2021). Finally, psychological scientific understanding may involve skills in imaginative or empathic reasoning (Eigner 2009), i.e., scientists’ ability to imagine being in their subjects’ position and envision their (normative) reasons for acting. This aligns with *Verstehen*, according to which psychological scientists should look for the perspective from which psychological phenomena appear ‘meaningful’ and ‘appropriate’. When psychological scientists can envision *why* people may act in specific ways, this can help them examine relevant consequences of their theories. Such imaginative or empathic reasoning, in turn, requires psychological scientists to have a level of awareness of the possible meaning that their subjects may ascribe to the concepts in their theories.

### Engineering sciences

One might object that engineering is not a science proper, and therefore not a suitable object for an account of *scientific* understanding. However, this objection underestimates the deep entanglement between science and engineering in contemporary research. In fields such as biomedical engineering, space science, materials science, and synthetic biology, the production of scientific knowledge and the design of technological artifacts are so thoroughly intertwined that drawing a clear boundary between "doing science" and "doing engineering" becomes impossible. Biomedical engineers, for instance, simultaneously investigate biological mechanisms and design medical devices, where the understanding gained from one activity directly informs the other. Similarly, space scientists develop novel instrumentation while studying astrophysical phenomena in such a way that their engineering work is inseparable from their scientific inquiry. Including engineering sciences in our comparative analysis is therefore not merely a matter of seeking "additional insights," but rather of recognizing that much contemporary scientific practice is also engineering practice. Examining how understanding operates in these hybrid contexts can illuminate both the reach and the limits of the contextual theory.

Engineering involves creating, improving, and fixing artefacts. Engineering science investigates the physico-chemical, structural, and relational details of artefacts thereby producing knowledge and understanding useful in engineering practice. This includes discovering novel material properties (e.g., conductivity of ceramics at high temperatures), explaining material failures (e.g., why metals break under load), or optimizing manufacturing procedures and system efficiency through computational methods. Although engineering spans diverse domains (civil, mechanical, nautical, software, etc.), each with unique aspects, certain explanatory approaches consistently recur throughout.

Engineering heavily relies on physical theories, and similarly, physical experiments depend extensively on engineered artifacts (Radder 2003). However, despite this interdependence, significant disciplinary differences exist. Take the example of early steam engines. Long before thermodynamics was formally understood, engineers had to develop models and explanations for the pressure buildup in boilers, material resistance, corrosion, decay, and other factors that weren't yet well understood (Boon 2008; Leveson 2012) to prevent ongoing explosions. Moreover, engineering often examines empirical regularities at abstraction levels beyond physics, where models present idealized abstractions that omit phenomena such as friction, chemical degradation, and material resistance. While theoretical corrections could account for these factors, such adjustments rarely yield empirical regularities that interest engineers (Boon 2008). What's often needed are different models, which combine higher-level features to realize functions.

Furthermore, unlike natural sciences that deal with existing phenomena, engineering often aims to create new artifacts and systems. Implementing novel artefacts and functions requires moving beyond representation to include projection (Vermaas and Vial 2018). In other words, engineering science not only seeks to understand actual phenomena, but also potential phenomena it can bring about and how these can be combined to achieve practical goals. Because of this, explanations are deeply intertwined with design models (Eckert and Hillerbrand 2018; see also Boon and Knuuttila 2009). A fundamental aspect of understanding in these design models involves functional decomposition: breaking down systems into functions, especially behavioural, effect, and purpose functions (van Eck and Weber 2014; 2017). Creating and applying these design models requires skills like analogical reasoning (Hey et al. 2018) and abductive reasoning (Koskela et al. 2018), in addition to the ability of functionally decompose systems.

The explanations that emerge primarily take mechanistic and causal forms (van Eck 2016; Barman and van Eck 2021; Barman 2022b). The degree of completeness and specificity of said mechanisms depends on the task at hand. In malfunction explanations, which are employed to understand why certain artefacts failed, or why certain systems do not function, engineers balance detailed descriptions of faulty components with broader abstraction of the overall mechanism (van Eck 2016). In reverse engineering and in redesign, where one tries to recreate or to improve an existing design, successful explanations accurately identify components that are constitutively relevant to a mechanism's operation, distinguishing them from merely causally influential factors (Van Eck 2018).

In this way, theoretical qualities in engineering explanations are tied to contextual explanatory aims. Take the case of accident causation models, which are used to understand accidents. The model chosen in each case is directly tied to the theoretical qualities it affords (Barman 2023). Systemic models, which account for complex, non-linear interactions, are most effective for improving safety due to their depth (i.e. the ability to account for a wide range of influencing factors and system-level constraints). In contrast, linear models are valued for their semantic simplicity, and are thus better suited for blame assignment by clearly tracing a single causal line. Epidemiological models, which treat accidents as the result of a combination of latent and active failures within a system, balance complexity with cognitive salience, thereby enabling presenting information in an accessible and intuitive format to lay audiences.

Notably, in the case of malfunction analysis, one can find a 'theoretical quality' that is arguably unique to engineering: "redesign utility", i.e. the ability to provide counterfactual information useful for improving design (Barman and van Eck 2021; Barman 2022a). While traditional qualities like simplicity, coherence, and depth are still relevant (Barman and van Eck 2021), the main feature of good explanations of failures is that they should contain information relevant towards improving the design; they should provide understanding (in the form of relations of counterfactual dependence at the right level of abstraction) as to how to redesign the system so as to avoid failure.

Lastly, given its practical nature, *intelligibility* is closely tied to practices aimed towards making models more intuitive or usable. This includes tools like CAD software, finite element analysis (FEA), control system models, and circuit simulators, which help engineers derive qualitative consequences without exact calculations. Though many of these techniques do involve detailed computation, their results can offer a qualitative picture. For example, an FEA output might visualize stress distribution in a building with a color scheme that highlights critical areas, allowing engineers to reason qualitatively about relevant structural properties.

### Mathematics

Philosophers of mathematics, including philosophers of mathematical practice, have discussed various types of understanding in the context of mathematics. These types include understanding proofs (Avigad 2008; Hamami and Morris 2024), understanding theorems (Inglis and Mejía-Ramos 2021), understanding definitions and understanding concepts or mathematical theories (Coumans, Frans and De Regt 2022).

However, the question is to what extent the contextual theory of understanding is applicable to mathematics. Starting with CUP, we observe that the term 'phenomenon' is not prevalent in the philosophy of mathematics. Mathematics can be said to be more abstract or theoretical in a sense. Phenomena could then be interpreted as statements, like theorems[4].

CUP also concerns ‘explanation of phenomenon P’. In the context of mathematics, this translates to ‘explanations of statement S’. In the philosophy of mathematics, explanations are often discussed in the context of explanatory proofs[5] even though the explanatory nature of other parts of mathematics has also been discussed.[6] Whereas all proofs show *that* a theorem is true, explanatory proofs also show *why* it is true. What this means has been discussed by e.g. Steiner (1978), Lange (2014), and Frans and Weber (2014). For instance, Steiner’s account associates explanatory proofs with proofs that relate the claim of a theorem to the characterizing property of the main object of that theorem. Although there are some technical issues with this account, it seems quite logical at first sight: if one wants to know why object O has a certain property Q, then deriving Q from the characterizing property of O would yield a satisfying answer.

Another hurdle in interpreting CUP in this context is the notion of theory and the role of models. One might argue that the ‘theory’ referred to in the contextual theory of understanding refers to the foundations, like Newton’s laws of motion, and not to all the theoretical results that can be derived from these laws. An analogue in the context of mathematics would constitute foundational axioms or definitions. Whereas definitions are prevalent (Coumans and Consoli 2023; Coumans 2023), foundational axioms do not really change in contemporary mathematical practice. This makes definitions, and the concepts they define, suitable interpretations of ‘theory’.

Lehet (2021a; 2021b) and Coumans, Frans and De Regt (2022) offer accounts of how definitions can contribute to understanding. Coumans, Frans and De Regt propose that definitions are accompanied by particular perspectives. They further argue that some perspectives are more useful for internalizing the body of knowledge and skills related to the concept that is defined[7] than others. For instance, seeing vectors in a vector space as linear combinations of basis vectors enables deriving a variety of properties related to vector spaces, like that vector spaces with the same finite dimension are isomorphic. A perspective being useful for internalizing knowledge and skills is a quality that might be interpreted as part of the intelligibility of a theory. Furthermore, models might be interpreted as intuitive arguments using these perspectives.

Another quality related to theories is naturalness. This can be described as grasping the concept’s properties that are responsible for a variety of results regarding this concept (Tappenden 2008). Another one is visualizability. Mathematicians indicate that they find it important to visualize concepts that they work with and prefer definitions that enable them to generate a mental image (Coumans and Consoli 2023). Finally, a skill that is related to mathematical understanding is mathematical intuition (D’Alessandro and Stevens 2024). Avigad (2008) describes understanding as selecting promising routes from the multitude of roads in the mathematical landscape. This falls under the D’Alessandro and Stevens’ (2024) description of mathematical intuition, which also covers being able to manipulate mathematical object in one’s mind, like visualizing dynamic moves in knot theory (de Toffoli and Giardino, 2014).

[4] Prima facie, it makes sense to look at Inglis and Mejía-Ramos (2021) for a theory of understanding theorems, but as this is discussed in an Understanding-first approach to explanations, its nature is different from the contextual theory of understanding as that is an Explanation-first approach.
[5] See D’Alessandro (2019) for an overview.
[6] Explanatory definitions are discussed by e.g. Lehet (2021a; 2021b) and Coumans, Frans and De Regt (2022).
[7] Internalizing the body of knowledge and skills is what Coumans, Frans and De Regt (2022), in line with Lehet (2021a, 2021b) and Inglis and Mejía-Ramos (2021), refer to as “factually understanding the definition”.

## 4. Conclusion: unity in diversity?

What conclusions can we draw if we compare the sketches provided in Section 3? To what extent is there agreement, or variation, in ways in which understanding is achieved across scientific disciplines? To begin with, it should be noted that two disciplines – engineering sciences and mathematics – are somewhat different in nature than the others, since achieving understanding of natural phenomena is not their main goal. Rather, mathematics aims at understanding abstract entities, in particular statements such as proofs, theorems and definitions. Still, it may contribute to scientific understanding and there may be interesting similarities between scientific and mathematical understanding. The engineering sciences, by contrast, aim first and foremost at understanding (the behavior of) artefacts. However, this typically involves understanding material properties of those artefacts and hence overlaps with e.g. physical and chemical understanding. So, including these disciplines in the comparison may provide additional insights.

Since we focus on explanatory understanding, let us start with considering the modes of explanation employed in the different disciplines. While there clearly is variation, it transpires that causal explanation is ubiquitous (with the obvious exception of mathematics). This is in line with the current consensus in the philosophy of science, and not surprising since causal explanation is a generic concept that can be articulated in different ways and hence is applicable in many sciences. In particular, mechanistic explanation can be viewed as a subclass of causal explanation, and is especially prominent in the 'special sciences' (geoscience, biology, neuroscience, psychology). Evolutionary explanations in biology complement mechanistic ones and can be seen as another subclass of causal explanation, focusing on ultimate rather than proximate causes. In several areas computer simulations are used to (re)construct causal or mechanistic accounts of the relevant phenomena (esp. in fields where these phenomena are not directly accessible, such as astrophysics or the geosciences). Non-causal varieties of explanation are more prominent outside the physical sciences. A prime example is functional explanation, which (not surprisingly) is used only in the life sciences, psychology, and in the engineering sciences. Neuroscience and psychology also witness an increase in mathematical explanation, especially dynamical and topological explanation. In general, the higher-level sciences, dealing with complex phenomena, are more pluralistic with regard to explanatory strategies than physics and chemistry. Remarkably, covering-law explanation is almost absent in our review: it is mentioned only in theoretical physics, confirming that the logical empiricist view on explanation was biased towards physics.

According to the contextual theory, understanding-providing explanations are typically model-based, where models mediate between theory and phenomena. Is this reflected in the descriptive accounts of explanatory understanding presented in Section 3? In particular, what is the nature and function of theories and models, and their relation in the different disciplines? While theories and models play a role in all scientific disciplines, the way in which they do varies greatly. First, in the physical sciences theories are typically mathematical and quantitative, and have a more prominent place in scientific practice than in the life sciences, where models occupy center stage. In theoretical physics and astrophysics fundamental theories such as quantum mechanics, quantum field theory, and general relativity are the starting point for model construction, where specific models of phenomena can be mathematical or pictorial (or both in case of computer simulations). This is in line with Morgan and Morrison's (1999) models-as-mediators thesis and with the contextual theory of scientific understanding. However, models can in turn also inspire further theory development. Fundamental physical theories, especially quantum mechanics, are relevant to chemistry as well, but in practice chemical understanding is usually achieved via qualitative chemical theories (which approximate the underlying quantitative physical theories) and associated models for specific phenomena. The same applies to the geosciences. While the view that geoscience is merely applied physics and chemistry is mistaken (Kleinhans, Buskes and De Regt 2005), it does indeed use theories from physics and chemistry (and sometimes biology), next to specific earth-scientific theories such as plate tectonics. Understanding of earth-scientific phenomena is achieved through the construction of models that range from computer simulations to analogue (e.g. scale) models.

In the higher-level sciences modeling takes precedence over theorizing, where models are typically less strongly connected to fundamental theories. Still, the latter remain important as a background for modeling practices, evolutionary theory being a prime example. Models in biology can be mechanistic or mathematical/computational. Evolutionary-developmental models are typically of the former kind, while optimization models exemplify the latter. Neuroscience is even more focused on modeling, while fundamental theories are almost absent. Mathematical, computational and network models are becoming increasingly important, although mechanistic models still dominate the scene (Kostić and Halffman 2023). A similar trend can be seen in psychology, where quantitative psychological science relies heavily on mathematical models for fitting empirical data. Such models have clear descriptive and predictive value but may lack in explanatory depth (and accordingly in providing understanding). Moreover, psychology features more general, qualitative theories, but these are not easily comparable with or testable by data models. Hence, there seems to be a theory-model gap, as in neuroscience. Finally, in the engineering sciences there is a key role for design models, which are informed by relevant disciplinary theories.[8]

So far, we can conclude that preferred explanatory strategies vary across scientific disciplines, although causal explanation in its various guises is a dominant mode. Explanation of phenomena generally involves model construction, where models can be of widely different kinds (corresponding to the different explanatory strategies employed). A particularly striking difference concerns the relation between models and theories in different disciplines: in the higher-level sciences theories play a subordinate role and modeling is often largely theory-independent, in contrast to physics, where models are tightly connected to fundamental theories. This has consequences for the applicability of the contextual theory of scientific understanding to scientific disciplines other than physics. In particular, it suggests that its central thesis that understanding requires intelligible theories is too restrictive – explanatory models may be constructed independently of theories. However, the intelligibility requirement itself is not misguided and there are good reasons to extend it to models. Hence the suggestion by Kostić and Khalifa (2025) to generalize CIT to CIR (where 'theory' is replaced by 'representation').[9] The corresponding revision of CUP would be:

> **CUP*:** A phenomenon *P* is understood scientifically if and only if there is an explanation of *P* that involves an intelligible representation R of P and conforms to the basic epistemic values of empirical adequacy and internal consistency.[10]

Granting that intelligibility applies to representations in general, which standards of intelligibility are employed in the various disciplines? Which properties of theories and models are valued for facilitating their use? Here it appears that there is variation but also agreement on the value of some properties. Especially visualizability is highly valued across the board, which aligns with the observed general preference for causal and mechanistic models/explanations. Another widely appreciated preference is for theories and models that can be interpreted in a qualitative way. Qualitative

[8] Mathematics does not feature theories and models in the sense discussed here, so we exclude this from the comparison.
[9] A change from theory intelligibility to model intelligibility was already suggested by Eigner (2010), in his analysis of the nature of psychological understanding.
[10] The definition of intelligibility should be adapted accordingly.

(interpretations of) theories and models afford intuitive insight into their consequences, in line with $CIT_1$/CIR. A more specific way to render theories and models intelligible is by invoking analogies, which can be used for both quantitative and qualitative interpretations. The skills that scientists need for constructing explanatory models vary in corresponding ways, ranging from abstract mathematical and computational skills, via analogical reasoning, to visual imagination and qualitative reasoning skills (where it should be noted that visual and intuitive reasoning is also practiced by mathematicians). Various disciplines feature their own specific skills, such as empathic reasoning in psychology and functional decomposition skills in the engineering sciences.

In conclusion, all scientific disciplines aim for understanding the phenomena in their domain, but this aim is achieved in a variety of ways: explanatory strategies, standards of intelligibility and criteria for understanding differ across and within disciplines. Still, there are common elements, most notably the crucial role of modeling. While the contextual theory of scientific understanding was developed to offer an account of understanding-providing practices in physics, it appears that it can also account for a variety of practices in other scientific disciplines. Even disciplines like biology and psychology, which prima facie differ greatly from physics and might seem to be incompatible with the contextual theory, can be accommodated by it at least in some respects and to some degree. Indeed, precisely because it acknowledges contextual variation in criteria and strategies for understanding, the contextual theory can be applied to a wide range of scientific disciplines. It can explain both the variation and the common core in the diversity of disciplinary understanding. However, it needs to be generalized in order to apply beyond the physical sciences: it should be modified such that it can accommodate disciplines that are primarily concerned with models instead of theories. We submit that criteria CUP* and CIR may serve as the basis for such a generalization.

Finally, what is the added value of a comparative philosophy of science approach for the study of scientific understanding? And what can aspiring comparative philosophers of science in turn learn from our comparative study of understanding? This paper is obviously just a first step in the direction of a substantial comparative analysis of scientific understanding. But it may give some clues as to what such an analysis would look like. First of all, while the disciplinary variation that we have documented seems to support pluralism about scientific understanding, the comparison between disciplines reveals a common core that would otherwise have gone unnoticed. Thus, for example, causal explanation, modeling, and visualization are elements that are shared across disciplines. As such, our cross-disciplinary examination suggests that understanding has a unity that any single-discipline investigation would have missed and that merits further systematic analysis. In this way, our practice-based, bottom-up approach inspires and contributes to a general philosophy of science treatment of understanding. And this is also why we chose to include the motto of the European Union in the title of our paper: *in varietate concordia* – united in diversity.

## Acknowledgments

- We thank two anonymous reviewers and the guest editors of this special issue for their helpful comments and suggestions.
- Kristian Barman's work was funded by Fonds Wetenschappelijk Onderzoek (grant number 1229124N).
- Milo Cornelissen acknowledges the *Studienstiftung des deutschen Volkes* for a doctoral scholarship.
- Paul Disberg acknowledges support from the Australian Research Council Centre of Excellence for Gravitational Wave Discovery (OzGrav) through project number CE230100016.
- Ludo Schoenmakers acknowledges the Konrad Lorenz Institute for Evolution and Cognition Research for support through a junior postdoctoral fellowship.